\documentclass[12pt]{amsart}
\usepackage{epsfig}
\begin{document}
\large
 \topmargin=5mm
 \oddsidemargin=-2mm
 \evensidemargin=-2mm
\baselineskip=24pt
\parindent 20pt
{\flushleft{\Large\bf  Binary Bell polynomials approach to the
integrability of nonisospectral and
  variable-coefficient nonlinear  equations }}\\[12pt]
{\large\bf  Engui Fan\footnote{ E-mail
address: \  faneg@fudan.edu.cn}  }\\[8pt]
{\small  School of Mathematical Sciences and Key Laboratory of
Mathematics for Nonlinear Science, Fudan University, Shanghai,
200433, P.R.  China} \vspace{6mm}
\begin{center}
\begin{minipage}{5.2in}
\baselineskip=20pt { \small {\bf Abstract.}   Recently, Lembert,
Gilson et al proposed a lucid and systematic approach to obtain
bilinear  B\"{a}cklund transformations and  Lax pairs  for
constant-coefficient soliton equations based on the use of binary
Bell polynomials. In this paper, we would like to further develop
this method with  new applications.  We extend this  method  to
systematically investigate complete integrability of nonisospectral
and  variable-coefficient equations.
 In addiction, a method is described for deriving  infinite conservation laws of nonlinear
evolution equations based on the use of  binary
 Bell polynomials.  All conserved density and flux are given  by
explicit recursion formulas. By taking variable-coefficient KdV and
KP equations as illustrative examples, their  bilinear formulism,
bilinear B\"{a}cklund transformations, Lax pairs, Darboux covariant
Lax pairs  and conservation laws are obtained in a quick and natural
manner. In conclusion, though the coefficient functions have
influences on a variable-coefficient nonlinear equation, under
certain constrains the equation  turn out to be also
  completely integrable, which leads us to a canonical interpretation of their $N$-soliton solutions
  in theory. 
}
\end{minipage}\\[20pt]
\end{center}
{\bf\large 1. Introduction}\\

 In many physical situations, it is often
preferable to have an equation with variable-coefficients, which may
allow us to describe real  phenomena in physical and engineering
fields. For example, variable-coefficient nonlinear
Schr¡§odinger-typed ones, which describe such situations more
realistically than their constant-coefficient counterparts, in
plasma physics, arterial mechanics and long-distance optical
communications \cite{Serkin}-\cite{Biswas}.  Many physical and
mechanical situations governed by variable-coefficient KdV (vc-KdV)
equation, e.g., the pulse wave propagation in blood vessels and
dynamics in the circulatory system, matter waves and nonlinear atom
optics enhanced by the observations of Bose-Einstein condensation in
the weakly interacting atomic gases, the nonlinear excitations of a
Bose gas of impenetrable bosons with longitudinal confinement, the
nonlinear waves in types of rods \cite{Zamir}-\cite{Dai}. In recent
years, there has been considerable interest in the study of
variable-coefficient nonlinear equations, such as vc-KdV, vc-KP,
vc-Schr\"{o}dinger,  vc-Boussinesq and cylindrical KdV equations.
Recent progress in the investigation of the complete integrability
and exact solutions for such equations via Painleve analysis,
inverse scattering transformation, Hirota bilinear method and
Darboux transformation has been reported, the details can be seen in
reference and references therein \cite{Chan1}-\cite{Gao}. It is
obvious that variable-coefficient equations are often more
complicated and  difficult to be solved than constant-coefficient
ones. As well-known, investigation of integrability for a nonlinear
equation can be regarded as a pre-test and  the first step of its
exact solvability. There are many significant properties, such as
Lax pairs, infinite conservation laws, infinite symmetries,
Hamiltonian structure, Painlev\'{e} test that can characterize
integrability of nonlinear equations. This may pave the way for
constructing their exact solutions explicitly in a future.  But in
contrast to constant-coefficient cases, very little of detail is
known about complete integrability of variable-coefficient nonlinear
equations.

 Among the direct algebraic methods applicable to nonlinear partial
 differential equations in soliton theory, there is on which has proved particularly
 powerful: the bilinear method developed by Hirota \cite{Hirota1, Hirota2}.
 Once a nonlinear equation is written in bilinear forms
by a dependent variable transformation, then multi-soliton solutions
are usually obtained  \cite{Hu}-\cite{Fan1}.
 The search for a Hirota representation of a given nonlinear equation is generally recognized as an
 important first step in the construction of multi-soliton solutions.
Yet, the construction of such bilinear B\"{a}cklund transformation
is not as one would wish. It relies on a particular skill in using
appropriate exchange formulas which are connected with the linear
representation of the system.  Recently, Lembert, Gilson et al
proposed an alternative procedure based on the use of Bell
polynomials which enable one to obtain parameter families of
bilinear B\"{a}cklund transformation for
 soliton equations in a lucid and systematic way
 \cite{Gilson}-\cite{Lambert2}. The Bell polynomials are found to
 play an important role in the characterization of bilinearizable
 equations. As a consequence bilinear B\"{a}cklund transformation with single field can be
 linearize into corresponding Lax pairs.
   Their method provides a shortest  way to bilinear  B\"{a}cklund transformation and Lax
pairs of nonlinear  equations,  which establishes a deep relation
between integrability of a nonlinear equation and the  Bell
polynomials.

 The problem that we consider in this paper is to further develop this method with
new applications.  We extend the binary Bell polynomials approach to
 a large class of  nonisospectral and variable-coefficient equations, such as nonisospectral and
 variable-coefficient KdV and KP equations etc. One of the many remarkable properties that deemed to characterize
soliton equations is existence of an infinite sequence of
conservation laws.   Here  we  propose a approach to construct
infinite conservation laws of nonlinear evolution equations through
decoupling binary Bell polynomials into a Riccati type equation and
a divergence type equation.  As illustrative examples, the  bilinear
representations, bilinear B\"{a}cklund transformations, Lax pais and
infinite conservation laws of the vc-KdV and vc-KP equations are
obtained in a quick and natural  manner.  The integrable constraint
conditions on the  variable-coefficient functions can be naturally
found in the procedure of applying binary Bell polynomials.
 We can also find that though the coefficient functions have influences on
  a variable-coefficient equation, under certain constrains the equation still can admit many
  integrability properties  which are similar to those of its standard
  constant-coefficient equation.
  The organization of this paper is as follows. In section 2, we briefly
  present necessary notations on multi-dimensional binary Bell polynomial
  that will be used in this paper.  In the sections 3 and 4, we deal with
integrability of nonisospectral vc-KdV equation and vc-KdV equation,
respectively. We aim at integrability of nonisospectral vc-KP
equation and vc-KP equation in the sections
5 and 6, respectively. \\[12pt]
{\bf\large  2.   Multi-dimensional binary Bell polynomials}\\

The main tool used in this paper is a class of the Bell polynomials,
named after E. T. Bell \cite{Bell}.  To make our presentation easy
understanding and self-contained, we simply  recall some necessary
notations  on the Bell polynomials, the details refer, for instance,
to Lembert and Gilson's work \cite{Gilson}-\cite{Lambert2}.

Let $f=f(x_1,\cdots, x_n)$ be a $C^{\infty}$ function with
multi-variables, the following polynomials
$$Y_{n_1x_1,\cdots,n_\ell x_\ell}(f)\equiv Y_{n_1,\cdots,n_\ell}(f_{r_1x_1,\cdots,r_\ell x_\ell})
=e^{-f}\partial_{x_1}^{n_1}\cdots\partial_{x_\ell}^{n_\ell} e^f$$ is
called multi-dimensional Bell polynomials, in which we denote that
$f_{r_1x_1,\cdots,r_\ell
x_\ell}=\partial_{x_1}^{r_1}\cdots\partial_{x_\ell}^{r_\ell}f,\ \
r_1=0,\cdots,n_1;\cdots; \ r_\ell=0,\cdots,n_\ell $.

For example,  for the simplest case  $f=f(x)$,  the associated
one-dimensional Bell polynomials read
$$\begin{aligned}
&{Y}_1(f) =f_x, \ {Y}_{2}(f)=f_{2x}+f_x^2,\ \ \
{Y}_3(f)=f_{3x}+3f_xf_{2x}+f_x^3, \cdots.
\end{aligned}$$
For $f=f(x,t)$,  the  of associated two-dimensional Bell polynomials
are
$$\begin{aligned}
&{Y}_{x,t}(f) =f_{x,t}+f_xf_t, \ \
{Y}_{2x,t}(f)=f_{2x,t}+f_{2x}f_t+2f_{x,t}f_x+f_x^2f_t, \cdots.
\end{aligned}$$

Based on the use of above Bell polynomials, the multi-dimensional
binary  Bell polynomials can be defined as follows
$$\mathcal{Y}_{n_1x_1,\cdots,n_{\ell}x_{\ell}}(v,w)=Y_{n_1,\cdots,n_{\ell}}
(f)\mid_{f_{r_1x_1,\cdots,r_{\ell}x_{\ell}}=\left\{\begin{matrix}v_{r_1x_1,
\cdots,r_{\ell}x_{\ell}},&r_1+\cdots+r_{\ell} \ \ {\rm is\ \
odd},\cr\cr w_{r_1x_1,\cdots,r_{\ell}x_{\ell}},&r_1+\cdots+r_{\ell}
\ \ {\rm is\ \ even},\end{matrix}\right.}$$ which inherit the easily
recognizable partial structure of the Bell polynomials.

 The first few lowest order binary Bell Polynomials are
 $$\begin{aligned}
&\mathcal{Y}_x (v)=v_x, \ \mathcal{Y}_{2x}(v,w)=w_{2x}+v_x^2,\ \ \mathcal{Y}_{x,t}(v,w)=w_{xt}+v_xv_t.\\
&\mathcal{Y}_{3x}=v_{3x}+3v_xw_{2x}+v_x^3, \cdots.
\end{aligned}\eqno(2.1)$$

The link between $\mathcal{Y}$-polynomials and the standard Hirota
bilinear equation $D_{x_1}^{n_1}\cdots D_{x_{\ell}}^{n_{\ell}}F\cdot
G$ can be given by an identity
$$\mathcal{Y}_{n_1x_1,\cdots,n_{\ell}x_{\ell}}(v=\ln F/G, w=\ln FG)=
(FG)^{-1}D_{x_1}^{n_1}\cdots D_{x_{\ell}}^{n_{\ell}} F\cdot
G,\eqno(2.2)$$ in which ${n_1}+n_2+\cdots+n_{\ell}\geq 1$.  In the
particular case when  $G=F$, the formula (2.2) becomes
 $$\begin{aligned}
&F^{-2}D_{x_1}^{n_1}\cdots D_{x_{\ell}}^{n_{\ell}} F\cdot
F=\mathcal{Y}_{n_1x_1,\cdots,n_{\ell}x_{\ell}}(0,q=2\ln
F)\\[6pt]
&=\left\{\begin{matrix}0,&n_1+\cdots+n_{\ell} \ \ {\rm is\ \
odd},\cr\cr
P_{n_1x_1,\cdots,n_{\ell}x_{\ell}}(q),&n_1+\cdots+n_{\ell} \ \ {\rm
is\ \ even},\end{matrix}\right.\end{aligned}\eqno(2.3)$$ in which
the $P$-polynomials can be  characterized by an equally recognizable
even part partitional structure
 $$\begin{aligned}
&P_{2x}(q)=q_{2x}, \ P_{x,t}(q)=q_{xt},  \ P_{4x}(q)=q_{4x}+3q_{2x}^2,\\
&P_{6x}(q)=q_{6x}+15q_{2x}q_{4x}+15q_{2x}^2, \cdots.
\end{aligned}\eqno(2.4)$$
The formulae (2.2) and (2.3) will prove particularly useful in
connecting nonlinear equations with their corresponding bilinear
equations. This means that once a nonlinear equation is expressible
as a linear combination of $P$-polynomials, then it  can be
transformed into a linear equation.

It follows that the binary Bell polynomials
$\mathcal{Y}_{n_1x_1,\cdots,n_{\ell}x_{\ell}}(v,w)$ can be separated
into $P$-polynomials and $Y$-polynomials
$$\begin{aligned}
&(FG)^{-1}D_{x_1}^{n_1}\cdots D_{x_{\ell}}^{n_{\ell}}F\cdot
G=\mathcal{Y}_{n_1x_1,\cdots,n_{\ell}x_{\ell}}(v, w)|_{v=\ln F/G, w=\ln FG}\\
&=\mathcal{Y}_{n_1x_1,\cdots,n_{\ell}x_{\ell}}(v, v+q,)|_{v=\ln F/G, q=2\ln G}\\
&=\sum_{n_1+\cdots+n_\ell=even}\sum_{r_1=0}^{n_1}\cdots\sum_{r_\ell=0}^{n_\ell}\prod_{i=1}^{\ell}
\left(\begin{matrix}n_i\cr
r_i\end{matrix}\right)P_{r_1x_1,\cdots,r_{\ell}x_{\ell}}(q)Y_{(n_1-r_1)x_1,\cdots,(n_\ell-r_\ell)x_\ell}(v).
\end{aligned}\eqno(2.5)$$
The key property of the multi-dimensional  Bell polynomials
$$Y_{n_1x_1,\cdots,n_{\ell}x_{\ell}}(v)|_{v=\ln\psi}=\frac{\psi_{n_1x_1,\cdots,n_{\ell}x_{\ell}}}{\psi},\eqno(2.6)$$
implies that the binary Bell polynomials
$\mathcal{Y}_{n_1x_1,\cdots,n_{\ell}x_{\ell}}(v,w)$ can still be
linearized by means of the Hopf-Cole transformation $v=\ln \psi$,
that is, $\psi=F/G$.  The formulae (2.5) and (2.6) will then provide
the shortest way to the associated Lax system of nonlinear
equations.

 We start with construction of infinite
conservation laws by virtue of binary Bell polynomials. We define a
new auxiliary field variable
$$\eta=(q'_{x_k}-q_{x_k})/2,$$
where $q'$ and $q$ are given by $q'=w+v, \ q=w-v$ and $x_k$ is a
appropriate  variable chosen from $x_1, \cdots, x_{\ell}$.  The
two-filed condition
$$C(q',q)=E(q')-E(q)=0\eqno(2.7)$$
can be regarded as the natural ansatz for a bilinear B\"{a}ckbend
transformation.  By expressing the two-filed condition (2.7) in
terms of binary Bell $\mathcal{Y}$-polynomials and their
derivatives,  we expect that the resulting condition is then
decoupled into a pair of constrains, i.e. often a Riccati type
equation with respect to $x_k$,
$$ \eta_{x_k}+f(\eta)=0,  \eqno(2.8)$$
and a divergence-type equation
$$ \partial_{x_1}F_1(\eta)+\cdots+\partial_{x_{\ell}}F_{\ell}(\eta)=0.\eqno(2.9)$$
The recursion formulas of  conversed density come from the equation
(2.8), the formulas of associated flux are obtained by  using the
equation (2.9). It is often the case that the first few conservation
laws of a nonlinear equation have a physical interpretation.
\\[12pt]
{\bf\large 3.   Nonisospectral variable coefficient KdV equation}\\

 Consider nonisospectral vc-KdV equation \cite{Chan1}
$$u_t+h_1(u_{3x}+6uu_x)+4h_2u_x-h_3(2u+xu_x)=0,\eqno(3.1)$$
where $h_1=h_1(t),\ h_2=h_2(t)$ and $\ h_3=h_3(t)$  are all
arbitrary functions with respect to time variable  $t$. The equation
(3.1) includes some governing  physical  equations as special
reduction, such  as celebrated constant-coefficient KdV equation
$$u_t+6uu_x+u_{3x}=0,$$
 cylindrical KdV equation ($h_1=1,\ h_2=1/8t,\
h_3=0$) \cite{Maxson}
$$u_t+6uu_x+u_{3x}+\frac{1}{2t}u_x=0,\eqno(3.2)$$
and the vc-KdV equation ($h_1=1,\ h_2=c_0/4,\ h_3=-\gamma$)
$$u_t+6uu_x+u_{3x}+\gamma u+[(c_0+\gamma x)u]_x=0,$$
  which describes the effect of relaxation inhomogeneous medium
\cite{Hirota3}.  It can be observed  that the equation (3.1) is
invariant under Galiean transformation
$$u\rightarrow u+\lambda, \ x\rightarrow x+6\lambda t, \ t\rightarrow t.
$$
 The inverse
scattering transformation  of the equation (3.1) was considered by
Chan and Li \cite{Chan1}. Lou and Ruan obtained infinite
conservation laws \cite{Lou}. Here we shall investigate the
integrability of the equation (3.1) from bilinear representation,
B\"{a}cklund transformation, Lax pair, Darboux covariant Lax pair
and infinite conservation laws.
\\[4pt]
{\bf 3.1.  Bilinear representation }

In order to detect its existence of linearizable representation, we
introduce a potential field $q$ by setting
$$u=c(t)q_{2x},\eqno(3.3)$$
with $c=c(t)$ being free function to be the appropriate choice such
that the equation (3.1) connect with $P$-polynomials.  Substituting
(3.3) into (3.1) and integrating with respect to $x$ yields
$$E(q)\equiv
q_{xt}+h_1(q_{4x}+3cq_{2x}^2)+4h_2q_{2x}-h_3(q_x+xq_{2x})+q_x\partial_t\ln
c=0.\eqno(3.4)$$ Comparing the second term of this equation together
with the formula (2.4), we require $c(t)=1$. The result equation is
then cast into a combination form of $P$-polynomials
$$E(q)=P_{xt}(q)+h_1P_{4x}(q)+4h_2P_{2x}(q)-h_3(xP_{2x}(q)+q_x)=0.\eqno(3.5)$$

 Making a change of dependent variable
$$q=2\ln F\ \ \Longleftrightarrow \ \ u=c q_{2x}=2(\ln F)_{2x}$$
and noting  the property (2.3),  the equation (3.5) gives the
bilinear representation  as follows
$$(D_xD_t+h_1D_x^4+4h_2D_x^2-xh_3D_x^2-h_3\partial_x)F\cdot
F=0,$$ in which we have used the notation   $\partial_x F\cdot
F\equiv\partial_x F^2=2FF_x$. This equation is easy to be solved for
multi-soliton solutions by using Hirota's bilinear method. For
example, the regular one-soliton like solution reads
 $$\begin{aligned}
&u=\frac{k^2}{2}{\rm sech}^2\frac{k x+\omega}{2},
\end{aligned}$$
where $k=k(t)$ and $\omega=\omega(t)$ are two functions about $t$,
given by
$$k(t)=\alpha e^{\int h_3 dt}, \ \
\omega(t)=-\int (h_1k^3+4h_2k)dt.$$ The multi-soliton solution are
omitted here since exactly solving the equation (3.1) is not our
main purpose in this paper.
\\[4pt]
{\bf 3.2.  B\"{a}cklund transformation and Lax pair}

 Next,  we search for the bilinear
B\"{a}cklund transformation and Lax pair of the vc-KdV equation
(3.1). Let $q$ and $q'$ be  two different solutions of the equation
(3.4), respectively,  we associate the two-field condition
 $$\begin{aligned}
&E(q')-E(q)=(q'-q)_{xt}+h_1[(q'-q)_{4x}+3(q'+q)_{2x}(q'-q)_{2x}]\\
&+4h_2(q'-q)_{2x}-h_3[(q'-q)_x+x(q'-q)_{2x}]=0.
\end{aligned}\eqno(3.5)$$
This two-field condition can be regarded as the natural ansatz for a
bilinear B\"{a}cklund transformation and  may produce the required
transformation under appropriate additional constraints.

To find such  constraints,  we  introduce two new variables
$$v=(q'-q)/2, \ \ w=(q'+q)/2,\eqno(3.6)$$
and   rewrite the condition (3.5) into the form
 $$\begin{aligned}
&E(q')-E(q)=v_{xt}+h_1(v_{4x}+6v_{2x}w_{2x})+4h_2v_{2x}-h_3(v_x+xv_{2x})\\
&=\partial_x[\mathcal{Y}_t(v)+h_1\mathcal{Y}_{3x}(v,w)]+R(v,w)=0,
\end{aligned}\eqno(3.7)$$
with
$$R(v,w)=3h_1{\rm Wronskian}[\mathcal{Y}_{2x}(v,w),
\mathcal{Y}_x(v)]+\partial_x[4h_2\mathcal{Y}_x(v)-xh_3\mathcal{Y}_x(v)].$$

In order to decouple  the two-field condition (3.7) into a pair of
constraints, we impose such a constraint which enable us to  express
$R(v,w)$ as the $x$-derivative of a  combination of
$\mathcal{Y}$-polynomials.  The simplest possible choice of such
constraint may be
$$\mathcal{Y}_{2x}(v,w)+\alpha\mathcal{Y}_x(v)=\lambda,\eqno(3.8)$$
where $\alpha$ and  $\lambda$  are arbitrary parameters.  On account
of the equation (3.8), then  $R(v,w)$ can be rewritten  in the form
$$R(v,w)=\partial_x[3h_1\lambda\mathcal{Y}_x(v)+4h_2\mathcal{Y}_x(v)-xh_3\mathcal{Y}_x(v)].\eqno(3.9)$$
Then from (3.7)-(3.9),  we deduce  a coupled
 system of $\mathcal{Y}$-polynomials
$$\begin{aligned}
&\mathcal{Y}_{2x}(v,w)+\alpha\mathcal{Y}_x(v)-\lambda=0,\\
&\partial_x\mathcal{Y}_t(v)+\partial_x[h_1\mathcal{Y}_{3x}(v,w)+
(3h_1\lambda+4h_2-xh_3)\mathcal{Y}_x(v)]=0.
\end{aligned}\eqno(3.10)$$
where prefer the second equation in the conserved form without
integration with respect to $x$, which is useful to construct
conservation laws later.  By application of the identity (2.2), the
system (3.10) immediately  leads to   the bilinear B\"{a}cklund
transformation
$$\begin{aligned}
&(D_x^2+\alpha D_x-\lambda)F\cdot G=0,\\
&[D_t+h_1D_x^3+ (3h_1\lambda+4h_2-xh_3)D_x+\beta]F\cdot G=0,
\end{aligned}$$
where $\beta$ is a arbitrary  parameter.

By transformation  $v=\ln \psi$, it follows  from  the formulae
(2.5) and (2.6) that
$$\begin{aligned}
&\mathcal{Y}_{x}(v)=\psi_{x}/\psi, \ \ \mathcal{Y}_{2x}(v,w)=q_{2x}+\psi_{2x}/\psi,\\
&\mathcal{Y}_{3x}(v,w)=3q_{2x}\psi_x/\psi+\psi_{3x}/\psi,\ \
\mathcal{Y}_t(v)=\psi_t/\psi,
\end{aligned}$$
on account of which,  the system (3.10) is then linearized into a
Lax pair with double parameters about $\lambda$ and $\beta$
$$\begin{aligned}
&L_1\psi\equiv(\partial_x^2+\alpha\partial_x+q_{2x})\psi=\lambda\psi,
\ \ \lambda_t=2h_3\lambda,
\end{aligned}\eqno(3.11)$$
$$\begin{aligned}
&(\partial_t+L_2)\psi\equiv
[\partial_t+h_1\partial_x^3+3h_1(q_{2x}+\lambda)\partial_x
+(4h_2-xh_3)\partial_x]\psi
\end{aligned}\eqno(3.12)$$
or equivalently replacing $q_{2x}$ by $u$,
$$\begin{aligned}
&\psi_{2x}+\alpha\psi_x+(u-\lambda)\psi=0, \ \ \lambda_t=2h_3\lambda, \\
&\psi_t+[h_1(2u+4\lambda+\alpha^2)+4h_2-xh_3]\psi_x
+[\beta-h_1u_x-\alpha h_1(u-\lambda)]\psi=0.
\end{aligned}$$
Starting from this Lax pair, the Darboux transformation and
soliton-like solutions  of the vc-KdV equation (3.1) can be
established \cite{Chan1}. It is easy to check that the integrability
condition
$$[L_1-\lambda, \partial_t+L_2]\psi=0$$
is satisfied if $u$ is a solution of  the vc-KdV equation (3.1) and
nonisospectral condition $\lambda_t=2h_3\lambda$ holds.
\\[4pt]
{\bf 3.4.  Infinite conservation laws}

 Finally, we  show
how to derive the infinite conservation laws for vc-KdV equation
(3.1) based on the use  of the binary Bell polynomials. The
conservation laws  actually have been hinted in  the two-filed
constraint system (3.10), which  can be rewritten  in the conserved
form
$$\begin{aligned}
&\mathcal{Y}_{2x}(v,w)+\alpha\mathcal{Y}_x(v)-\lambda=0,\\
&\partial_t\mathcal{Y}_x(v)+\partial_x[h_1\mathcal{Y}_{3x}(v,w)+
(3h_1\lambda+4h_2-xh_3)\mathcal{Y}_x(v)]=0.
\end{aligned}\eqno(3.13)$$
by applying the relation
$\partial_x\mathcal{Y}_t(v)=\partial_t\mathcal{Y}_x(v)=v_{xt}.$

 By introducing a new potential function
  $$\eta=(q'_x-q_x)/2, $$
it follows from the relation (3.6) that
$$v_x=\eta, \ \ w_x=q_x+\eta.\eqno(3.14)$$
Substituting (3.14) into (3.13), we get a Riccati-type equation
 $$\begin{aligned}
&\eta_x+\eta^2+q_{2x}=\lambda=\varepsilon^2,
\end{aligned}\eqno(3.15)$$
and a divergence-type equation
$$\begin{aligned}
&\eta_t+\partial_x[h_1\eta_{2x}+6h_1(\eta+\varepsilon)\varepsilon^2-2h_1(\eta+\varepsilon)^3
+ (4h_2-xh_3)(\eta+\varepsilon)]=0,
\end{aligned}\eqno(3.16)$$
where we have used the equation (3.15) to get the equation (3.16)
and set $\lambda=\varepsilon^2$.

To proceed, inserting the expansion
$$\eta=\varepsilon+\sum_{n=1}^{\infty} I_n(q,
q_x,\cdots)\varepsilon^{-n},\eqno(3.17)$$
 into the equation (3.15)
and  equating the coefficients for power of $\varepsilon$, we then
obtain  the recursion relations for $I_n$
$$\begin{aligned}
&I_1=-p_x=-\frac{1}{2}u,\ \ \ I_2=\frac{1}{4}p_{2x}=\frac{1}{4}u_x,\\
 &I_{n+1}=-\frac{1}{2}(I_{n,x}+\sum_{k=1}^{n}I_k I_{n-k}), \ \ n=2,
3, \cdots,
\end{aligned}\eqno(3.18)$$

By applying  the nonisospectral condition
$$\lambda_t=2h_3\lambda\ \ \Longrightarrow \varepsilon_t=h_3\varepsilon,$$
then substituting (3.17) into (3.16) yields
$$\begin{aligned}
&\sum_{n=1}^{\infty}
I_{n,t}\varepsilon^{-n}+\partial_x\left[h_1\sum_{n=1}^{\infty}I_{n,2x}
\varepsilon^{-n}-6h_1\varepsilon(\sum_{n=1}^{\infty}I_{n}
\varepsilon^{-n})^2
-2h_1(\sum_{n=1}^{\infty}I_{n}\varepsilon^{-n})^3\right.\\
&\left.+(4h_2-xh_3)\sum_{n=1}^{\infty}I_{n}\varepsilon^{-n}
-h_3\sum_{n=1}^{\infty}n\partial_x^{-1}I_{n}\varepsilon^{-n}\right]=0,
\end{aligned}$$
which provides us  infinite consequence of  conservation laws
$$I_{n,t}+F_{n,x}=0, \ n=1, 2, \cdots.\eqno(3.19)$$
In the equation (3.19), the conversed densities $I_n's$ are given by
formula (3.15) and the fluxes $F_n's$ are given by recursion
formulas explicitly
$$\begin{aligned}
&F_1=-\frac{1}{2}\left[h_1(u_{2x}+3u^2)+4h_2u-h_3(xu+\partial^{-1}_x u)\right],\\
&F_2=\frac{1}{4}\left[h_1(u_{3x}+6uu_x)+4h_2u_{x}-h_3(xu_{x}+2u)\right],\\
 &F_{n}=h_1I_{n,2x}-6h_1\sum_{k=1}^{n}I_k I_{n+1-k}-2h_1\sum_{i+j+k=n}I_iI_j I_{k}+
 (4h_2-xh_3)I_n\\
 &\ \ \ \ \ \ \
 \ \ +nh_3\partial_x^{-1}I_n, \ \ n=3, 4, \cdots.
\end{aligned}\eqno(3.20)$$
We present recursion formulas for generating an infinite sequence of
conservation laws for each equation, the first few conserved density
and associated flux  are explicit. The first equation of
conservation law equation (3.19) is exactly  the vc-KdV equation
(3.1). The expressions (3.20) indicate  that the fluxes $F_n's$ of
the vc-KdV equation are not local, which are different from  those
of standard constant-coefficient KdV equation.
 In conclusion, the vc-KdV equation (3.1) is
complete integrable in the sense that it admits bilinear
B\"{a}cklund transformation, Lax pair and infinite conservation
laws.
\\[12pt]
{\bf\large 4.  Generalized variable-coefficient KdV equation}\\

A more general example, we consider vc-KdV equation \cite{Tian}
$$u_t+h_1u_{3x}+h_2uu_x+h_3u_x+h_4u=0,\eqno(4.1)$$
where $h_j=h_j(t),\ j=1, 2, 3, 4$ are all arbitrary functions with
respect to time variable  $t$. Special cases of the equation (4.1)
include cylindrical equation \cite{Hlavaty}-\cite{Fan}
$$u_t+f(t)uu_x+g(t)u_{3x}=0$$
 and
 other special variable-coefficient equation \cite{Grimshaw, Joshi}
$$u_t+at^nuu_x+bt^mu_{3x}=0.$$
Recently,  Zhang et al obtained  the bilinear form, B\"{a}cklund
transformation and exact solutions  for the equation (4.1) under the
constrain \cite{Tian}
$$h_1=c_0 h_2e^{-\int h_4dt}.\eqno(4.2)$$
Here we construct bilinear representation, B\"{a}cklund
transformation, Lax pair and conservation laws of the equation (4.1)
based on the sue of binary  Bell polynomials technique, which will
be seen to be a  natural way to find such a constraint (4.2).
 We find that the bilinear representation of the
equation (4.1) existence without need of any constraint. The
constraint only need it for construction of the B\"{a}cklund
transformation,  Lax pair and conservation laws.
\\[4pt]
{\bf 4.1.  Bilinear representation}

 As before, we introduce a field $q$ by setting
$$u=c(t)q_{2x},\eqno(4.3)$$
in which $c=c(t)$ is free function  to be determined. Substituting
(4.3) into (4.1) and integrating with respect to $x$  yields
$$E(q)\equiv q_{xt}+h_1q_{4x}+\frac{1}{2}h_2cq_{2x}^2+h_3q_{2x}+(h_4+\partial_t\ln c)q_x=0.\eqno(4.4)$$
which can be cast into  a combination form of $P$-polynomials
 by using the formula (2.4)
$$E(q)=P_{xt}(q)+h_1P_{4x}(q)+h_3P_{2x}(q)+(h_4+\partial_t\ln h_1h_2^{-1})q_x=0,\eqno(4.5)$$
if one chooses the function  $c(t)=6h_1h_2^{-1}.$

By  transformation
$$q=2\ln F\ \ \Longleftrightarrow \ \  u=c(t)q_{2x}=12h_1h_2^{-1}(\ln F)_{2x}$$
and using the property (2.3), then the equation (4.5)  implies  the
bilinear form for the vc-KdV equation (4.1) as follows
$$[D_xD_t+h_1D_x^4+h_3D_x^2+(h_4+\partial_t\ln h_1h_2^{-1})\partial_x]F\cdot
F=0,\eqno(4.6)$$ which is obviously more general than that obtained
in \cite{Tian}, since we have no any constraint on the $h_1, h_2,
h_3$ and
 $h_4$. Starting the bilinear equation (4.6), we can get  multi-soliton
solutions to the vc-KdV equation (4.1). For example, one-soliton
solution takes the form
$$\begin{aligned}
&u=6h_1h_2^{-1}k^2{\rm sech}^2\frac{kx+\omega(t)}{2},
\end{aligned}$$
where $k$ is a constant and $\omega(t)$ given by
$$\omega(t)=-\int(k^3h_1+kh_3+h_4+\partial_t\ln h_1h_2^{-1})dt.$$
\\[4pt]
{\bf 4.2.  B\"{a}cklund transformation and Lax pair}

 In
order to obtain the bilinear B\"{a}cklund transformation and Lax
pairs of the equation (4.1),  let $q$, $q'$ be two solutions of the
equation (4.4) and consider  the associated  two-field condition
 $$\begin{aligned}
&E(q')-E(q)=(q'-q)_{xt}+h_1[(q'-q)_{4x}+3(q'+q)_{2x}(q'-q)_{2x})\\
&+h_3(q'-q)_{2x}+(h_4+\partial_t\ln h_1h_2^{-1})(q'-q)_x]=0,
\end{aligned}\eqno(4.7)$$
which may produce the required bilinear B\"{a}cklund transformation
under an appropriate additional constraint. By introducing variables
$$v=(q'-q)/2, \ \ w=(q'+q)/2\eqno(4.8)$$
we can  rewrite the condition (4.7) as the form
 $$\begin{aligned}
&E(q')-E(q)=v_{xt}+h_1(v_{4x}+6v_{2x}w_{2x})+h_3v_{2x}+(h_4+\partial_t\ln h_1h_2^{-1})v_x\\
&=\partial_x[\mathcal{Y}_t(v)+h_1\mathcal{Y}_{3x}(v,w)]+R(v,w)=0,
\end{aligned}\eqno(4.9)$$
with
$$R(v,w)=3h_1{\rm Wronskian}[\mathcal{Y}_{2x}(v,w),
\mathcal{Y}_x(v)]+h_3v_{2x}+(h_4+\partial_t\ln h_1h_2^{-1})v_x.$$ In
order to express $R(v,w)$ as the $x$-derivative of a linear
combination of $\mathcal{Y}$-polynomials, we choose a constraint
$$\mathcal{Y}_{2x}(v,w)+\alpha\mathcal{Y}_x(v)=\lambda,\eqno(4.10)$$
where $\alpha$ and $\lambda$ are arbitrary parameters.  Direct
calculation gives
$$R(v,w)=3h_1\lambda v_{2x}+h_3v_{2x}+(h_4+\partial_t\ln h_1h_2^{-1})v_x,$$
which can be written as $x$-derivative of $\mathcal{Y}$-polynomials
$$R(v,w)=\partial_x[3h_1\lambda\mathcal{Y}_x(v)+h_3\mathcal{Y}_x(v)+(h_4+\partial_t\ln h_1h_2^{-1})v].\eqno(4.11)$$
From (4.10)-(4.11), we infer that
$$\begin{aligned}
&\mathcal{Y}_{2x}(v,w)+\alpha\mathcal{Y}_x(v)-\lambda=0,\\
&\partial_x\mathcal{Y}_t(v)+\partial_x[h_1\mathcal{Y}_{3x}(v,w)+
(3h_1\lambda+h_3)\mathcal{Y}_x(v)+(h_4+\partial_t\ln
h_1h_2^{-1})v]=0,
\end{aligned}\eqno(4.12)$$
which can be cast into a bilinear B\"{a}cklund transformation by
using the property (2.2)
$$\begin{aligned}
&(D_x^2+\alpha D_x-\lambda)F\cdot G=0,\\
&[D_t+h_1D_x^3+ (3h_1\lambda+h_3)D_x+\beta]F\cdot G=0,
\end{aligned}\eqno(4.13)$$
if we set  the constraint
$$\begin{aligned}h_4+\partial_t\ln h_1h_2^{-1}=0 \ \ \Longrightarrow h_1=c_0e^{-\int h_4dt}h_2,\end{aligned}$$
with $c_0$ being a arbitrary parameter. Without loss of generality,
taking $c_0=1/6$, then
$$c(t)=6h_1/h_2=e^{-\int h_4dt}.$$  As $\alpha=\beta=0$, the B\"{a}cklund
transformation (4.13) reduces to the one obtained in  \cite{Tian}.

Making use of the Hopf-Cole  transformation  $v=\ln \psi$ and the
formula (2.5)-(2.6), then the system (4.10) can be linearized into a
Lax pair
$$\begin{aligned}
&L_1\psi=(\partial_x^2+\alpha\partial_x+q_{2x})\psi=\lambda\psi,
\end{aligned}\eqno(4.14)$$
$$\begin{aligned}
&(\partial_t+L_2)\psi=[\partial_t+h_1\partial_x^3+(3h_1q_{2x}+3\lambda
h_1+h_3)\partial_x]\psi=0,
\end{aligned}\eqno(4.15)$$
  or equivalently,
$$\begin{aligned}
&\psi_{2x}+\alpha\psi_x+(ue^{\int h_4dt}-\lambda)\psi=0, \\
&\psi_t+[h_1(2ue^{\int
h_4dt}+4\lambda+\alpha^2)+h_3]\psi_x+[\beta+\alpha\lambda-h_1(u_xe^{\int
h_4dt}-\alpha ue^{\int h_4dt})]\psi=0.
\end{aligned}$$
This Lax pair  can be used to construct Darboux transformation,
inverse scattering transformation  for soliton solutions. It is easy
to check that the integrability condition
$$[L_1-\lambda, \partial_t+L_2]\psi=0$$
is satisfied if $u$ is a solution of  the vc-KdV equation (4.1).
\\[4pt]
{\bf 4.3. Darboux covariant Lax pair}

Let us go back to the vc-KdV equation (4.1)  and the associated Lax
pair (4.14)-(4.15).  Assume that $\phi$ is a solution eigenvalue
equation (4.14) (taking $\alpha=0$ for simplicity). It is well-known
that the gauge transformation
$$T=\phi\partial_x\phi^{-1}=\partial_x-\sigma, \ \
\sigma=\partial_x\ln \phi\eqno(4.16)$$ map the operator
$L_1=\partial_x^2+q_{2x}$ onto a similar operator:
$$T(L_1(q)-\lambda)T^{-1}=\tilde{{L}}_1(\tilde{q})-\lambda,$$
which satisfies the covariance
condition
$$\tilde{{L}}_1(\tilde{q})=L_1(\tilde{q}=q+\Delta q),\  \ {\rm
with} \ \ \Delta q=2\ln\phi.$$ But it can verified that  similar
property does not hold for the evolution equation  (4.15).  Next
step is to find another third order operator ${L}_{2,{\rm cov}}(q)$
with appropriate coefficients, such that $\partial_t+L_{2,{\rm
cov}}(q)$ be mapped, by gauge transformation (4.16), onto a similar
operator $\partial_t+\tilde{L}_{2,{\rm {\rm cov}}}(\tilde{q})$ which
satisfies the {\rm cov}ariance condition
$$\tilde{L}_{2,{\rm cov}}(\tilde{q})={L}_{2,{\rm cov}}(\tilde{q}=q+\Delta q).$$

Suppose that $\phi$ is a solution of the following Lax pair
$$\begin{aligned}
&L_1\phi=\lambda\phi,\\
 &(\partial_t+{L}_{2,{\rm cov}})\phi=0, \ \
{L}_{2,{\rm cov}}=4h_1\partial_x^3+b_1\partial_x+b_2,
\end{aligned}\eqno(4.17)$$
where $b_1$ and $b_2$ are functions to be determined. It suffice
that we require that the transformation $T$ map the operator
$\partial_t+L_{2,{\rm cov}}$ onto the similar one
$$T(\partial_t+L_{2,{\rm cov}})T^{-1}=\partial_t+\tilde{L}_{2,{\rm cov}}, \ \
\tilde{L}_{2,{\rm
cov}}=4h_1\partial_x^3+\tilde{b}_1\partial_x+\tilde{b}_2,\eqno(4.18)$$
where $\tilde{b}_1$ and $\tilde{b}_2$ satisfy the {\rm cov}ariant
condition
$$\tilde{b}_j={b}_j(q)+\Delta b_j={b}_j(q+\Delta q), \ \ j=1, 2.\eqno(4.19)$$

It follows from (4.16) and (4,18) that
$$\Delta b_1=\tilde{b}_1-b_1=12h_1\sigma_x, \ \ \Delta b_2=\tilde{b}_2-b_2=b_{1,x}+\sigma\Delta
b_1 +12h_1\sigma_{2x},\eqno(4.20)$$ and $\sigma$ satisfies
$$\sigma_t+4h_1\sigma_{1x}+b_{2,x}+\Delta
b_2\sigma+\tilde{b}_1\sigma_x=0.\eqno(4.21)$$ According to (4.19),
it remains to determine $b_1$ and $b_2$ in the form of polynomial
expressions in terms of derivatives of $q$
$$b_j=F_j(q,q_{x}, q_{2x}, q_{3x}, \cdots), \ \ j= 1, 2$$
such that
$$\Delta F_j=F_j(q+\Delta q, q_{x}+\Delta q_{x}, q_{2x}+\Delta q_{2x}, \cdots)
-F_j(q, q_{x},q_{2x},  \cdots) =\Delta b_j,\eqno(4.22)$$ with
$\Delta q_{rx}=2(\ln q)_{rx}, \ r=1, 2, \cdots$, the $\Delta b_j$
being determined by the relations (4.19).

Expanding the left hand of the equation (4.22) , we obtain
$$\Delta b_1=\Delta F_1=F_{1,q}\Delta q+F_{1,q_{x}}\Delta q_{x}+F_{1,q_{2x}}\Delta q_{2x}+ \cdots
=12h_1\sigma_x=6h_1\Delta q_{2x},$$ which implies that we can choose
$$b_1=F_1(q_{2x})=6h_1 q_{2x}+c_1(t),\eqno(4.23)$$
with $c_1(t)$ being arbitrary function about $t$.

From the eigenvalue equation in (4.17), we can find the following
relation
$$q_{3x}=-\sigma_{2x}-2\sigma\sigma_x.\eqno(4.24) $$
Substituting (4.23) and (4.24) into (4.19) leads to
$$\Delta b_2=12h_1q_{3x}+12h_1\sigma\sigma_x+12h_1\sigma_{2x}=6h_1\sigma_{2x}=3h_1\Delta q_{3x}.$$
The second condition
$$\Delta F_2=F_{2,q}\Delta q+F_{2,q_{x}}\Delta q_{x}+F_{2,q_{2x}}\Delta q_{2x}+
F_{2,q_{3x}}\Delta q_{3x}+ \cdots=\Delta b_2,$$ can be satisfied, if
one chooses
$$b_2=F_2(q,q_{x}q_{2x},q_{3x})=3h_1 q_{3x}+c_2(t),$$
in which $c_2(t)$ is arbitrary constant.

Setting $c_1(t)=h_3, \ c_2(t)=0$, we find the following Darboux {\rm
cov}ariant evolution equation
$$(\partial_t+L_{2,{\rm cov}})\phi=0, \ \
L_{2,{\rm cov}}=4h_1\partial_x^3+(6h_1q_{2x}+h_3)\partial_x+3h_1
q_{3x},$$
 which is in agreement with  the equation (4.21).
 Moreover, the relation between the operator $L_{2,{\rm cov}}$ and the operator $L_2$ is given by
$$L_{2,{\rm cov}}=L_2+3h_1\partial_x(L_1-\lambda).$$

The integrability condition of  the Darboux covariant Lax pair
(4.15) precisely give rise to the  equation (4.1) in Lax
representation
$$[\partial_t+L_{2,{\rm cov}}, L_1]=-
[q_{xt}+h_1q_{4x}+3h_1q_{2x}^2+h_3q_{2x}+h_4q_x ]_x.$$ The  higher
operators can be obtained in a similar way  step by step
$$L_{k,{\rm cov}}(q)=4h_1\partial_x^k+b_1\partial_x^{k-2}+\cdots+b_p,\ \ k=3, 4, \cdots$$
which are Darboux {\rm cov}ariant with respect to $L_1$, so as to
produce higher order members of the vc-KdV hierarchy.
\\[4pt]
 {\bf 4.4.  Infinite conservation laws}

Finally, we  construct the conservation laws of vc-KdV equation. The
second equation of (4.12) has been  conserved form due to the
relation
$\partial_x\mathcal{Y}_t(v)=\partial_t\mathcal{Y}_x(v)=v_{xt}.$
 By introducing a new potential function
  $$\eta=(q'_x-q_x)/2, $$
 it follows from the relation (4.8) that
$$v_x=\eta, \ \ w_x=q_x+\eta.\eqno(4.25)$$
Substituting (4.25) into (4.12), we get  a Riccati type equation
 $$\begin{aligned}
&\eta_x+\eta^2+q_{2x}=\lambda=\varepsilon^2,
\end{aligned}\eqno(4.26)$$
and a divergence type equation
$$\begin{aligned}
&\eta_t+\partial_x[h_1\eta_{2x}+6h_1(\eta+\varepsilon)\varepsilon^2-2h_1(\eta+\varepsilon)^3
+ h_3(\eta+\varepsilon)]=0,
\end{aligned}\eqno(4.27)$$
where we have used the equation (4.26) to get the equation (4.27)
and set $\lambda=\varepsilon^2$.

Substituting  the expansion
$$\eta=\varepsilon+\sum_{n=1}^{\infty} I_n(q,
q_x,\cdots)\varepsilon^{-n}.\eqno(4.28)$$ into the equation (4.26)
and equating the coefficients of $\varepsilon^{-1}$, the conserved
densities are explicitly  obtained  by  recursion relations
$$\begin{aligned}
&I_1=-p_x=-\frac{1}{2}e^{\int h_4dt}u,\ \ I_2=\frac{1}{4}e^{\int h_4dt}u_{x},\\
 &I_{n+1}=-\frac{1}{2}(I_{n,x}+\sum_{k=1}^{n}I_k I_{n-k}), \ \ n=2,
3, \cdots,
\end{aligned}\eqno(4.29)$$

In addition, substituting (4.28) into (4.27) leads to
$$\begin{aligned}
&\sum_{n=1}^{\infty}
I_{n,t}\varepsilon^{-n}+\partial_x\left[h_1\sum_{n=1}^{\infty}I_{n,2x}\varepsilon^{-n}
-6h_1\varepsilon(\sum_{n=1}^{\infty}I_{n} \varepsilon^{-n})^2
-2h_1(\sum_{n=1}^{\infty}I_{n}\varepsilon^{-n})^3+h_3\sum_{n=1}^{\infty}I_{n}\varepsilon^{-n}\right]=0,
\end{aligned}$$
which provides us  infinite conservation laws
$$I_{n,t}+F_{n,x}=0, \ n=1, 2, \cdots.\eqno(4.30)$$
In the equation (4.30), the conversed densities $I_n's$ are given by
recursion formulas (4.29),  and the fluxes $F_n's$ are given by
$$\begin{aligned}
&F_1=-\frac{1}{2}e^{\int h_4dt}(h_1u_{2x}+3h_2u^2+h_3u),\\
&F_2=\frac{1}{4}e^{\int h_4dt}(h_1u_{3x}+6h_2e^{\int h_4dt}uu_x+h_3u_x),\\
 &F_{n}=h_1I_{n,2x}-6h_1\sum_{k=1}^{n}I_k I_{n+1-k}-2h_1\sum_{i+j+k=n}I_iI_j I_{k}+
 h_3I_n, \ \ n=3, 4, \cdots.
\end{aligned}$$
The first equation of (4.30) is exactly the vc-KdV equation (4.1).
We see that above fluxes $F_n's$ can be obtained from solution $u$
by algebraic and differential manipulation, thus they are local.
Taking the boundary condition of $u$ into account, the conservation
equation (4.30) implies that $\{I_n, \ n=1,2, \cdots,\}$ constitute
infinite conserved densities of the vc-KdV equation (4.1). In
conclusion, the vc-KdV equation (4.1) is  complete integrable under
the constraint $h_2=6h_1e^{\int h_7dt}$ in the sense that it admits
bilinear B\"{a}cklund transformation, Lax pair and infinite
conservation laws.
\\[12pt]
{\bf\large 5.   Nonisospectral  variable-coefficient KP equation}\\

 Consider nonisospectral and vc-KP equation \cite{Chan3}
 $$\begin{aligned}
&u_t+h_1(u_{3x}+6uu_x+3\alpha^2\partial_x^{-1}u_{yy})+h_2(u_x-\alpha
xu_y-2\alpha\partial_x^{-1}u_y)\\
&-h_3(xu_x+2u+2yu_y)=0,
\end{aligned}\eqno(5.1)$$
 where $h_1=h_1(t),\ h_2=h_2(t)$
and $\ h_3=h_3(t)$  are all arbitrary functions with respect to time
variable  $t$. The equation (5.1) reduces to the vc-KdV equation
(3.1) when  $u=u(x,t)$ is independent of the variable $y$. In the
case $h_1=1, \ h_2=h_3=0$, it reduce to standard KP equation.
\\[4pt]
{\bf 5.1.  Bilinear representation}

 By
introducing a potential  field $q$
$$u=c(t) q_{2x},\eqno(5.2)$$
with $c=c(t)$ is a free function about $t$ to be determined, the
resulting equation (5.1) for $q$ (integrating with respect to $x$)
reads
$$\begin{aligned}
&E(q)\equiv q_{xt}+h_1(q_{4x}+3cq_{2x}^2+3\alpha^2q_{2y})
+h_2(q_{2x}-x\alpha q_{xy}-\alpha
q_y)\\
&-h_3(q_x+xq_{2x}+2yq_{xy})+q_x\partial_t\ln
c=0,\end{aligned}\eqno(5.3)$$ which can be expressed in the form of
$P$-polynomials
 $$\begin{aligned}
&E(q)=P_{xt}(q)+h_1[P_{4x}(q)+3\alpha^2P_{2y}(q)]+h_2[P_{2x}(q)-\alpha x P_{xy}(q)-\alpha q_y]\\
&-h_3[xP_{2x}(q)+2yP_{xy}(q)+q_x]=0,
\end{aligned}\eqno(5.4)$$
if one chooses $c(t)=1$ and use the formula (2.4).

By application of the variable transformation
$$q=2\ln F\ \ \Longleftrightarrow \ \  u=2(\ln F)_{2x}$$
and using the property (2.3), then the equation (5.4)  give  the
bilinear form for the vc-KP equation (5.1) as follows
$$[D_xD_t+h_1(D_x^4+3\alpha^2 D_y^2)+h_2(D_x^2-\alpha x D_xD_y-\alpha\partial_y)
-h_3(xD_x^2+2yD_xD_y+\partial_x)]F\cdot F=0,$$ starting from which,
we can get multi-soliton solutions to the equation (5.1). For
example, one-soliton solution takes the form
$$\begin{aligned}
&u=\frac{(k+s)^2}{2}{\rm sech}^2\frac{\xi+\zeta+\ln\omega}{2},
\end{aligned}$$
in which $\xi=k x-k^2y/\alpha,\ \zeta=s x+s^2y/\alpha$, while
$k=k(t), s=s(t)$ and  $\omega=\omega(t)$ are all functions about
$t$, satisfying
$$\begin{aligned}
&k_t=h_3k-bk^2, \ \ s_t=h_3 s+bs^2,\\
&\omega(t)=\exp(\int [h_3(k^3+s^3)-h_2(k+s)]dt).
\end{aligned}$$
\\[4pt]
{\bf 5.2.  B\"{a}cklund transformation and Lax pair}

 In
following we consider bilinear B\"{a}cklund transformation and Lax
pairs of the equation (5.1).  Assume $q$ and $q'$ are two solutions
of the equation (5.3),  we consider  the corresponding  two-field
condition
 $$\begin{aligned}
&E(q')-E(q)=(q'-q)_{xt}+h_1[(q'-q)_{4x}+3(q'+q)_{2x}(q'-q)_{2x}+3\alpha^2(q'-q)_{2y}]\\
&+h_2[(q'-q)_{2x}-\alpha x(q'-q)_{xy}-\alpha(q'-q)_y]-h_3[x(q'-q)_{2x}\\
&+2y(q'-q)_{xy}+(q'-q)_x]=0,
\end{aligned}\eqno(5.5)$$
which may produce the required bilinear B\"{a}cklund transformation
under an appropriate additional constraint.

By change  of the variables
$$v=(q'-q)/2, \ \ w=(q'+q)/2\eqno(5.6)$$
we rewrite the condition (5.5) in the form
 $$\begin{aligned}
&E(q')-E(q)=v_{xt}+h_1(v_{4x}+6v_{2x}w_{2x}+3\alpha^2v_{2y})+h_2(v_{2x}-\alpha
x v_{xy}-\alpha
v_y)\\
&-h_3(xv_{2x}+2yv_{xy}+v_x)=\partial_x[\mathcal{Y}_t(v)+h_1\mathcal{Y}_{3x}(v,w)]+R(v,w)=0,
\end{aligned}\eqno(5.7)$$
with
$$\begin{aligned}
&R(v,w)=3h_1{\rm Wronskian}[\mathcal{Y}_{2x}(v,w),
\mathcal{Y}_x(v)]+3h_1\alpha^2v_{2y}+h_2(v_{2x}-\alpha x
v_{xy}-\alpha v_y)\\
&-h_3(xv_{2x}+2yv_{xy}+v_x).\end{aligned}$$ In order to express
$R(v,w)$ as the $x$-derivative of a linear combination of
$\mathcal{Y}$-polynomials, we choose the constraint
$$\mathcal{Y}_{2x}(v,w)+\alpha\mathcal{Y}_y(v)=\lambda,\eqno(5.8)$$
on account of which
$$\begin{aligned}
&R(v,w)=\partial_x\{h_1[3\lambda \mathcal{Y}_{x}(v)-3\alpha
\mathcal{Y}_{xy}(v,w)] +h_2[\mathcal{Y}_{x}(v)-\alpha x
\mathcal{Y}_{y}(v)]\\
&-h_3[x\mathcal{Y}_{x}(v)+2y\mathcal{Y}_{y}(v)]\}.\end{aligned}\eqno(5.9)$$

Combining relations (5.7)-(5.9), we then obtain  a pair of
constraints in Bell polynomial form
$$\begin{aligned}
&\mathcal{Y}_{2x}(v,w)+\alpha\mathcal{Y}_y(v)-\lambda=0,\\
&\partial_t\mathcal{Y}_{x}(v)+\partial_x\{h_1[\mathcal{Y}_{3x}(v,w)-3\alpha
\mathcal{Y}_{xy}(v,w)+3\lambda \mathcal{Y}_{x}(v)]
+h_2[\mathcal{Y}_{x}(v)-\alpha x
\mathcal{Y}_{y}(v)]\\
&-h_3[x\mathcal{Y}_{x}(v)+2y\mathcal{Y}_{y}(v)]\}=0,
\end{aligned}\eqno(5.10)$$
which result in  bilinear B\"{a}cklund transformation
$$\begin{aligned}
&(D_x^2+\alpha D_y-\lambda)F\cdot G=0,\\
&[D_t+h_1(D_x^3-3\alpha D_xD_y+3\lambda D_x)+h_2(D_x-\alpha x
D_y)\\
&-h_3(xD_x+2yD_y)+\beta]F\cdot G=0
\end{aligned}\eqno(5.11)$$
by applying the property (2.2).

It only remains to linearize the expression (5.10) with the formulae
(2.5) and (2.6).  By using the Hopf-Cole transformation $v=\ln
\psi$, the bilinear B\"{a}cklund transformation (5.10) can be
linearized into a Lax pair
$$\begin{aligned}
&(\partial_y+L_1)\psi\equiv \psi_y+\alpha^{-1}\psi_{2x}+\alpha^{-1}(q_{2x}-\lambda)\psi=0, \ \ \lambda_t=2h_3\lambda, \\
&(\partial_t+L_2)\psi\equiv \psi_t+4h_1\psi_{3x}+(xh_2+2y\alpha^{-1}h_3)\psi_{2x}+(6h_1q_{2x}+h_2-xh_3)\psi_x\\
&\ \ \ \ \ \ \ \ \ \ \ \ \ \ \ \ +[3h_1q_{3x}-3\alpha
h_1q_{xy}+(q_{2x}-\lambda)(xh_2+2y\alpha^{-1}h_3)]\psi=0,
\end{aligned}$$
which can be used to construct Darboux transformation, inverse
scattering transformation  for getting  soliton solutions. It is
easy to check that the integrability condition
$$[\partial_y+L_1, \partial_t+L_2]\psi=0$$
is satisfied if $u$ is a solution of  the vc-KP equation (5.1) and
nonisospectral condition $\lambda_t=2h_3\lambda$
holds.\\[4pt]
{\bf 5.3.  Infinite conservation laws}

Finally, we precede to construct the conservation laws of vc-KP
equation. For this purpose, we decompose the two-filed condition
(5.7) into $x$- and $y$-derivative of a linear combination of
$\mathcal{Y}$-polynomials. Let us go back to the two-field condition
(5.7) and consider the following decomposition
$$\begin{aligned}
&R(v,w)=\partial_x[3h_1\lambda v_x-3\alpha h_1v_xv_y+h_2(v_x-\alpha
xv_y)-h_3(xv_x+2yv_y)]\\
&+\partial_y(3h_1\alpha^2 v_y+3h_1\alpha
v_x^2).\end{aligned}\eqno(5.11)$$ It follows from (5.7), (5.8) and
(5.11) that
$$\begin{aligned}
&\mathcal{Y}_{2x}(v,w)+\alpha\mathcal{Y}_y(v)-\lambda=0,\\
&(\mathcal{Y}_{x}(v))_t+\partial_x\{h_1[\mathcal{Y}_{3x}(v,w)-3\alpha
\mathcal{Y}_{y}(v,w)+3\lambda \mathcal{Y}_{x}(v)]
+h_2[\mathcal{Y}_{x}(v)-\alpha x
\mathcal{Y}_{y}(v)]\\
&-h_3[x\mathcal{Y}_{x}(v)+2y\mathcal{Y}_{y}(v)]\}+\partial_y[3h_1\alpha^2
\mathcal{Y}_{y}(v)+3h_1\alpha\mathcal{Y}_{x}^2(v)]=0,
\end{aligned}\eqno(5.12)$$
which is slightly different from (5.10) and can produce desired
conservation laws.
 By introducing a new potential function
  $$\eta=(q'_x-q_x)/2, $$
and it follows from the relation (5.6) that
$$v_x=\eta, \ \ w_x=q_x+\eta.\eqno(5.13)$$

Substituting (5.13) into (5.12), we decompose the two-field
condition (5.7) into  a Riccati type equation
 $$\begin{aligned}
&\eta_x+\eta^2+\alpha\partial^{-1}_x\eta_y+q_{2x}-\varepsilon^2=0,
\end{aligned}\eqno(5.14)$$
and a divergence type equation
$$\begin{aligned}
&\eta_t+\partial_x[h_1(\eta_{2x}+6\eta\varepsilon^2-2\eta^3-6\alpha
\eta\partial_x^{-1}\eta_y )+h_2(\eta-x\alpha
\partial_x^{-1}\eta_y)\\
&-h_3(x\eta+2y\partial_x^{-1}\eta_y)]+\partial_y(3h_1\alpha^2\partial_x^{-1}\eta_y+3h_1
\alpha \eta^2)=0,
\end{aligned}\eqno(5.15)$$
where we have used the equation (5.14) to get the equation (5.15)
and set $\lambda=\varepsilon^2$.

Substituting the expansion
$$\eta=\varepsilon+\sum_{n=1}^{\infty} I_n(q,
q_x,\cdots)\varepsilon^{-n}.\eqno(5.16)$$ Inserting the equation
(5.16) into the equation (5.14), equation the coefficients for power
of $\varepsilon$, then we have the recursion relations for $I_n$
$$\begin{aligned}
&I_1=-\frac{1}{2}q_{2x}=-\frac{1}{2}u,\ \ I_2=
\frac{1}{4}(u_{2x}+\alpha \partial^{-1}_x u_{y}),\ \
I_3=-\frac{1}{8}(u_{3x}+u^2+\alpha u_{y}),\\
 &I_{n+1}=-\frac{1}{2}(I_{n,x}+\sum_{k=1}^{n}I_k
I_{n-k}+\alpha\partial^{-1}_x I_{n,y}), \ \ n=3, 4, \cdots,
\end{aligned}\eqno(5.17)$$
Again substituting (5.16) into (5.15) and noting the nonisospectral
condition
$$\lambda_t=2h_3\lambda\ \ \Longrightarrow \varepsilon_t=h_3\varepsilon,$$
we then obtain the following infinite conservation laws
$$I_{n,t}+F_{n,x}+G_{n,y}=0, \ n=1, 2, \cdots.\eqno(5.18)$$
In the equation (5.18), the conversed densities $I_n's$ are given by
formula (5.17), the first fluxes $F_n's$ are given by
$$\begin{aligned}
&F_1=h_1 I_{1,2x}+(h_2-xh_3)I_1-(x\alpha h_1+2y h_3)\partial^{-1}_x
I_{1,y}-6\alpha h_1
\partial^{-1}_x I_{2,y}+h_3\partial^{-1}_x I_1,\\
&F_2=h_1 I_{2,2x}+(h_2-xh_3)I_2-(x\alpha h_1+2y h_3)\partial^{-1}_x
I_{2,y} -12h_1I_1I_2-12\alpha h_1 I_1\partial^{-1}_x I_{2,y}\\
&-6\alpha h_1
\partial^{-1}_x I_{3,y}+2h_3\partial^{-1}_x I_2,\\
 &F_{n}=-6h_1\sum_{k=1}^{n}I_k (I_{n+1-k}+\alpha \partial_x^{-1} I_{n-k,y})
 -2h_1\sum_{i+j+k=n}I_iI_j I_{k}+
 (h_2-xh_3)I_n+h_1I_{n,2x}\\
 &\ \ \ \ \ \ \
 \ \ -(x\alpha h_2+2yh_3)\partial_x^{-1} I_{n,y} +nh_3\partial_x^{-1} I_{n}
 -6\alpha h_1 I_{n+1}, \ \ n=3, 4, \cdots.
\end{aligned}$$
and the second fluxes $G_n's$ are given by
$$\begin{aligned}
&G_1=3h_1\alpha^2\partial^{-1}_xI_{1,y},\ \
G_2=3h_1\alpha I_1^2+3h_1\alpha^2\partial^{-1}_xI_{2,y},\\
&G_{n}=3h_1\alpha\sum_{k=1}^{n}I_k
I_{n-k}+3h_1\alpha^2\partial_x^{-1}I_{n,y}, \ \ n=2, 3, \cdots.
\end{aligned}$$

The first equation of (5.18) is exactly the vc-KP equation (5.1).
The expressions $F_n's$ and $G_n's$  indicate that the fluxes of the
vc-KP equation are not local. To summarize, the vc-KP equation (5.1)
is complete integrable, since it admits  bilinear B\"{a}cklund
transformation, Lax pair and infinite conservation laws.
\\[12pt]
{\bf\large 6.   General  variable-coefficient KP equation}\\

 Consider a general vc-KP equation \cite{Liu}
$$(u_t+h_1u_{3x}+h_2uu_x)_x+h_3u_{2y}+h_4u_{xy}+(h_5+h_6y)u_{2x}
+h_7u_x=0,\eqno(6.1)$$
 where $h_i=h_i(t), \ i=1, 2, \cdots, 7$  are  arbitrary functions with respect to time
variable  $t$.  The equation (6.1) include many  special
variable-coefficient equations  in physics, such as cylindrical KdV
equation \cite{Hlavaty}
$$u_t+uu_x+u_{3x}+\frac{1}{2t}u_x=0,\eqno(6.2)$$
cylindrical KP equation \cite{Nakamura, Johnson}
$$(u_t+h_1u_{3x}+h_2uu_x)_x+\frac{1}{2t}u_x+\frac{3\sigma^2}{t^2}u_{2y}=0,\eqno(6.3)$$
generalized cylindrical KP equation \cite{David, Levi}
$$(u_t+h_1u_{3x}+h_2uu_x)_x+\frac{1}{2t}u_x+\frac{3\sigma^2}{t^2}u_{2y}+r(t)u_{xy}
+[f(t)+g(t)y]u_{2x}=0.\eqno(6.4)$$ Here we attempt to find the
integrability condition that  the equation (6.1) possesses bilinear
representation, B\"{a}ckbend transformation,  Lax pair, Darboux
covariant Lax pair and infinite conservation laws.
\\[4pt]
{\bf 6.1.  Bilinear representation}

By introducing a potential field $q$
$$u=c(t) q_{2x},$$
with $c=c(t)$ is free function to be determined, the resulting
equation (6.1) for $q$ (integrating with respect to $x$ twice) reads
$$\begin{aligned}
&E(q)\equiv q_{xt}+h_1q_{4x}+\frac{c}{2}h_2q_{2x}^2+h_3q_{2y}+
h_4q_{xy}+(h_5+h_6y)q_{2x}\\
&+(h_7+\partial_t\ln c)q_x=0,\end{aligned}\eqno(6.5)$$ which can be
expressible as $P$-polynomials
 $$\begin{aligned}
&E(q)=P_{xt}(q)+h_1P_{4x}(q)+h_3P_{2y}(q)+h_4P_{xy}(q)+(h_5+h_6)P_{2x}(q)\\
&+(h_7+\partial_t\ln h_1h_2^{-1})q_x=0,
\end{aligned}\eqno(6.6)$$
if one chooses $c=6h_1h_2^{-1}$ and use the formula (2.4). By
application of the transformation
$$q=2\ln F\ \ \Longleftrightarrow \ \  u=cq_{2x}=12h_1h_2^{-1}(\ln F)_{2x}$$
and using the property (2.3), then the equation (6.6)  gives  the
bilinear representation for the vc-KP equation (6.1) as follows
$$[D_xD_t+h_1D_x^4+h_3 D_y^2+h_4D_xD_y+(h_5+h_6y)D_x^2+(h_7+\partial_t
\ln h_1h_2^{-1})\partial_x]F\cdot F=0.$$ Starting form this bilinear
equation, we can get multi-solutions, for example, the regular
one-soliton like solution is
$$\begin{aligned}
&u=6h_1h_2^{-1}k^2{\rm sech}^2\frac{kx+sy+\omega}{2},
\end{aligned}$$
in which $k$ is a arbitrary constant, while  $s=s(t)$ and
$\omega=\omega(t)$ are two function with respect to $t$,  given by
 $$ s(t)=k\int h_6dt, \ \ \omega(t)=-\int(k^3h_1+k^{-1}s^2h_3+sh_4+h_7+\partial_t\ln
h_1h_2^{-1})dt.$$
\\[4pt]
{\bf 6.2.   B\"{a}cklund transformation and Lax pair}

We now consider bilinear B\"{a}cklund transformation and Lax pairs
of the equation (6.1). Let $q'$ and $q$ be two solutions of  the
equation (6.5),  we consider the following two-field condition
 $$\begin{aligned}
&E(q')-E(q)=(q'-q)_{xt}+h_1(q'-q)_{4x}+3h_1(q'+q)_{2x}(q'-q)_{2x}+h_3(q'-q)_{2y}\\
&+h_4(q'-q)_{xy}+(h_5+h_6y)(q'-q)_{2x}+(h_7+\partial_t\ln
h_1h_2^{-1})(q'-q)_x=0,
\end{aligned}\eqno(6.7)$$
which may produce the required bilinear B\"{a}cklund transformation
under an appropriate additional constraint.

On introducing two new  variables
$$v=(q'-q)/2, \ \ w=(q'+q)/2\eqno(6.8)$$
we rewrite the condition (6.7) as the form
 $$\begin{aligned}
&E(q')-E(q)=v_{xt}+h_1v_{4x}+6h_1v_{2x}w_{2x}+h_3v_{2y}+h_4v_{xy}+(h_5+h_6y)v_{2x}\\
&+(h_7+\partial_t\ln h_1h_2^{-1})v_x
=\partial_x[\mathcal{Y}_t(v)+h_1\mathcal{Y}_{3x}(v,w)]+R(v,w)=0,
\end{aligned}\eqno(6.9)$$
with
$$\begin{aligned}
&R(v,w)=3h_1{\rm Wronskian}[\mathcal{Y}_{2x}(v,w),
\mathcal{Y}_x(v)]+h_3v_{2y}+h_4v_{xy}+(h_5+h_6y)v_{2x}\\
&+(h_7+\partial_t\ln h_1h_2^{-1})v_x.\end{aligned}$$

In order to express $R(v,w)$ as the $x$- and $y$-derivative of
$\mathcal{Y}$-polynomials, we choose the constraint
$$\mathcal{Y}_y(v)+\alpha(t)\mathcal{Y}_{2x}(v,w)=\lambda,\eqno(6.10)$$
where $\alpha=\alpha(t)$ is to be determined. Direct calculation
show that
$$\begin{aligned} &R(v,w)=3h_1\lambda
v_{2x}-\alpha^{-1}[h_3w_{2x,y}+(2h_3-3h_1\alpha^2)v_xv_{xy}+3h_1\alpha^2v_{2x}v_y]\\
&+h_4v_{xy}+(h_5+h_6y)v_{2x} +(h_7+\partial_t\ln
h_1h_2^{-1})v_x,\end{aligned}\eqno(6.11)$$ which can be expressible
$R(v,w)$ as the $x$-derivative of a linear combination of
$\mathcal{Y}$-polynomials
$$\begin{aligned} &R(v,w)=\partial_x[3h_1\lambda
\mathcal{Y}_{x}-3\alpha
h_1\mathcal{Y}_{3x}(v,w)+h_4\mathcal{Y}_{y}+(h_5+h_6y)\mathcal{Y}_{x}],\end{aligned}$$
if we take a simple constraints
$$h_7+\partial_t\ln
h_1h_2^{-1}=0, \ \ h_3=2h_3-3h_1\alpha^2=3h_1\alpha^2,$$ namely,
$$h_2=6h_1e^{\int h_7dt}, \ \ \ 3\alpha^2=h_3h_1^{-1}.$$
From (6.9)-(6.11), one infers that
$$\begin{aligned}
&\mathcal{Y}_{2x}(v,w)+\alpha\mathcal{Y}_y(v)-\lambda=0,\\
&\partial_x\mathcal{Y}_t(v)+\partial_x\{h_1[\mathcal{Y}_{3x}(v,w)-3\alpha
\mathcal{Y}_{xy}(v,w)+3\lambda \mathcal{Y}_{x}(v)]
+h_4\mathcal{Y}_{y}\\
&+(h_5+h_6y)\mathcal{Y}_{x}\}=0,
\end{aligned}\eqno(6.12)$$
which leads  to bilinear B\"{a}cklund transformation of variable
coefficient KP equation
$$\begin{aligned}
&(D_x^2+\alpha D_y-\lambda)F\cdot G=0,\\
&[D_t+h_1(D_x^3-3\alpha D_xD_y+3\lambda D_x)+h_4D_y+(h_5+h_6y)
D_x+\beta]F\cdot G=0
\end{aligned}$$
by using the property (2.2).

By using the Hopf-Cole transformation  $v=\ln \psi$ and  the
formulae (2.5) and (2.6), the system (6.12) can be linearized into a
Lax pair
$$\begin{aligned}
&(\alpha\partial_y+L_1)\psi\equiv(\alpha\partial_y+\partial_x^2+q_{2x}-\lambda)\psi=0,  \\
&(\partial_t+L_2)\psi\equiv [\partial_t+4h_1\partial_x^3-h_4\alpha^{-1}\partial_x^2+(6h_1q_{2x}+h_5+h_6y)\partial_x\\
&\ \ \ \ \ \ \ \ \ \ \ \ \ \ \  +3h_1q_{3x}-3 h_1\alpha
q_{xy}-h_4\alpha^{-1}q_{2x}+h_4\alpha^{-1}\lambda]\psi=0,
\end{aligned}\eqno(6.13)$$
 whose integrability condition
$$[\alpha\partial_y+L_1, \partial_t+L_2]\psi=0$$
is satisfied if $u$ is a solution of  the vc-KP equation (6.1) and
$\alpha_t=0$, or equivalently, $h_3h_1^{-1}={\rm constant}$. This
Lax pair can be used to construct Darboux transformation,
inverse scattering transformation  for soliton solutions to  the vc-KP equation (6.1).\\[4pt]
{\bf 6.3. Darboux  covariant Lax pair}

Let us go back to the vc-KP equation (6.1)  and the associated Lax
pair (6.13).  Assume that $\phi$ is a solution of the following Lax
pair
$$\begin{aligned}
&(\alpha\partial_y+L_1)\phi=\lambda\phi,\ \
L_1=\partial_x^2+q_{2x},\\
 &(\partial_t+{L}_{2,{\rm cov}})\phi=0, \ \
{L}_{2,{\rm cov}}=h_1\partial_x^3+b_1\partial_x^2+b_2\partial_x+b_3,
\end{aligned}\eqno(6.14)$$
where $b_1$, $b_2$ and $b_3$ are functions to be determined. It is
shown that the gauge transformation
$$T=\phi\partial_x\phi^{-1}=\partial_x-\sigma, \ \
\sigma=\partial_x\ln \phi\eqno(6.15)$$ map the operator
$\alpha\partial_y+L_1(q)$ onto a similar operator:
$$T(\alpha\partial_y+L_1(q))T^{-1}=\alpha\partial_y+\tilde{{L}}_1(\tilde{q}=q+\Delta q)\  \ {\rm
with} \ \ \Delta q=2\ln\phi.$$ It is suffices to verify that such
transformation (6.15) map the $\partial_t+{L}_{2,{\rm cov}}$ into
similar one
$$T(\partial_t+L_{2,{\rm cov}})T^{-1}=\partial_t+\tilde{L}_{2,{\rm cov}}, \ \
\tilde{L}_{2,{\rm
cov}}=h_1\partial_x^3+\tilde{b}_1\partial_x^2+\tilde{b}_2\partial_x+\tilde{b}_3,\eqno(6.16)$$
where $\tilde{b}_j, j=1, 2, 3$ and  $\tilde{L}_{2,{\rm cov}}$
satisfy the covariant conditions
$$\tilde{b}_j={b}_j(q)+\Delta b_j={b}_j(q+\Delta q), \ \ j=1, 2, 3.$$
$$\tilde{L}_{2,{\rm cov}}(\tilde{q})={L}_{2,{\rm cov}}(\tilde{q}=q+\Delta q), \ \ \Delta
q=2\ln\phi.$$

It follows from (6.15) and (3,16) that
$$\begin{aligned}
&\Delta b_1=\tilde{b}_1-b_1=0,\ \
 \Delta b_2=12h_1\sigma_x+b_{1,x}+\Delta b_1\sigma, \\
 &\Delta b_3=12h_1\sigma_{2x}+2\sigma_x\tilde{b}_{1}+\Delta
b_2\sigma+b_{2,x},
\end{aligned}\eqno(6.17)$$
 We require  $b_1, b_2$ and $b_3$ in the  differential polynomial form of potential filed $q$
$$b_j=F_j(q,q_{x}, q_{2x}, q_{3x}, \cdots), \ \ j= 1, 2$$
such that
$$\Delta F_j=F_j(q+\Delta q, q_{x}+\Delta q_{x}, q_{2x}+\Delta q_{2x}, \cdots)
-F_j(q, q_{x},q_{2x},  \cdots) =\Delta b_j,$$ with $\Delta
q_{rx}=2(\ln q)_{rx}, \ r=1, 2, \cdots$.

From eigenvalue equation in (6.14), we get the relation
$$q_{3x}=-\alpha \sigma_{xy}-(\sigma_x+\sigma^2)_x,$$
on account of which, solving the system (6.17) yields
$$\begin{aligned}
&b_1=c_1(y,t), \ \ b_2=6h_1q_{2x}+c_2(y,t),\\
&b_3=3h_1(q_{3x}-\alpha q_{xy})+c_1q_{2x}+c_3(y,t),
\end{aligned}\eqno(6.18)$$ with $c_1(y,t), c_2(y,t)$ and $c_3(y,t)$ being
arbitrary functions with respect to $y$ and $t$.  From (6.14) and
(6.18),  we then find a  Darboux {\rm cov}ariant evolution equation
$$
\begin{aligned}
&(\partial_t+L_{2,{\rm cov}})\phi=0, \\
&L_{2,{\rm
cov}}=4h_1\partial_x^3+c_1\partial_x^2+(6h_1q_{2x}+c_2)\partial_x+
3h_1q_{3x}-3h_1\alpha q_{2x}+c_3.\end{aligned}$$
 In particular, if
setting
$$c_1=-h_4\alpha^{-1}, \ c_2=h_5+h_6y, \ c_3=h_4\lambda,$$
 the Darboux covariant operator $L_{2,{\rm cov}}$ reduce to operator
$L_2$, namely $$L_{2,{\rm cov}}=L_2.$$
 Therefore the Lax pair (6.13)
is Darboux covariant under constraint $h_3h_1^{-1}={\rm constant}$.

In a similar way, we can get higher  operators
$$L_{p,{\rm cov}}(q)=h_1\partial_x^p+b_1\partial_x^{p-2}+\cdots+b_p,\ \ p=3, 4, \cdots$$
which are Darboux  covariant with respect to $L_1$ step by step, so
as to produce higher order members of the vc-KP hierarchy.
 \\[4pt]
{\bf 6.4.  Infinite conservation laws}

Finally, we turn to construct the conservation laws of vc-KP
equation. For this purpose, we expect re-decompose the two-filed
condition (6.7) into $x$- and $y$-derivative of
$\mathcal{Y}$-polynomials. We return to revisit   $R(v,w)$ in
two-field condition (6.9) and write it  as another form
$$\begin{aligned}
&R(v,w)=[3h_1\lambda v_x-3\alpha
h_1v_xv_y+(h_5+h_6y)v_{x}]_x+(h_3v_{y}+h_4v_{x}+3h_1\alpha
v_x^2)_y.\end{aligned}\eqno(6.19)$$ It follows from the relations
(6.9), (6.10) and (6.19) that
$$\begin{aligned}
&\mathcal{Y}_{2x}(v,w)+\alpha\mathcal{Y}_y(v)-\lambda=0,\\
&\partial_t\mathcal{Y}_{x}(v)+\partial_x[h_1\mathcal{Y}_{3x}(v,w)-3h_1\alpha
\mathcal{Y}_{y}(v,w)+3h_1\lambda \mathcal{Y}_{x}(v)
+(h_5+h_6y)\mathcal{Y}_{x}(v)]\\
&+\partial_y[h_3\mathcal{Y}_{y}(v)+3h_1\alpha
\mathcal{Y}_{x}(v)^2+h_4\mathcal{Y}_{x}(v)]=0,
\end{aligned}\eqno(6.20)$$
which is slightly different from (6.10) and can produce desired
conservation laws. Especially we don't need the constraint
$3\alpha^2=h_3h_1^{-1}$.

 By introducing a new potential function
  $$\eta=(q'_x-q_x)/2, $$
and it follows from the relation (6.8) that
$$v_x=\eta, \ \ w_x=q_x+\eta.\eqno(6.21)$$
Substituting (6.21) into (6.20) yields a coupled system
$$\begin{aligned}
&\eta_x+\eta^2+\alpha\partial^{-1}_x\eta_y+q_{2x}-\varepsilon^2=0,
\end{aligned}\eqno(6.22)$$
$$\begin{aligned}
&\eta_t+\partial_x[h_1(\eta_{2x}+6\eta\varepsilon^2-2\eta^3-6\alpha
\eta\partial_x^{-1}\eta_y
)+(h_5+h_6y)\eta]\\
&+\partial_y(h_3\partial_x^{-1}\eta_y+3h_1 \alpha \eta^2+h_4\eta)=0,
\end{aligned}\eqno(6.23)$$
where we have used the equation (6.22) to get the equation (6.23)
and set $\lambda=\varepsilon^2$.

Substituting the expansion
$$\eta=\varepsilon+\sum_{n=1}^{\infty} I_n(p,
p_x,\cdots)\varepsilon^{-n}\eqno(6.24)$$  into the equation (6.22),
equating  the coefficients for power of $\varepsilon$, then we
obtain the recursion relations for $I_n$ as follows
$$\begin{aligned}
&I_1=-\frac{1}{2}q_{2x}=-\frac{1}{2}ue^{\int h_4dt},\ \ I_2=
\frac{1}{4}e^{\int h_4dt}(u_{2x}+\alpha \partial^{-1}_x u_{y}),\\
 &I_{n}=-\frac{1}{2}(I_{n,x}+\sum_{k=1}^{n}I_k
I_{n-k}+\alpha\partial^{-1}_x I_{n,y}), \ \ n=2, 3, \cdots,
\end{aligned}\eqno(6.25)$$
Again substituting (6.24) into (6.23) and  comparing  the
coefficients for power of $\varepsilon$ provide  us  infinite
conservation laws
$$I_{n,t}+F_{n,x}+G_{n,y}=0, \ n=1, 2, \cdots.\eqno(6.26)$$
In the equation (6.26), the conversed densities $I_n's$ obtained by
recursion formulas (6.25),  and the first fluxes $F_n's$ are
expressible in $I_n's$
$$\begin{aligned}
&F_1=h_1 I_{1,2x}-6\alpha h_1 I_1^2-6\alpha h_1
\partial^{-1}_x I_{2,y}+(h_5+h_6y)I_1,\\
&F_2=h_1 I_{2,2x}-6h_1I_1(2I_2+\alpha I_1+\alpha\partial^{-1}_x
I_{1,y})-6h_1\alpha \partial^{-1}_x I_{3,y}+(h_5+h_6y)I_2,\\
 &F_{n}=-6h_1\sum_{k=1}^{n}I_k (I_{n+1-k}+\alpha \partial_x^{-1} I_{n-k,y})
 -2h_1\sum_{i+j+k=n}I_iI_j I_{k}+h_1I_{n,2x}\\
 &\ \ \ \ \ \ \
 \ \ -6\alpha h_1 \partial^{-1}_xI_{n+1,y}+(h_5+h_6y) I_{n}
 , \ \ n=3, 4, \cdots.
\end{aligned}$$
and the second fluxes $G_n's$ are given by
$$\begin{aligned}
&G_1=h_3\partial^{-1}_xI_{1,y}+6\alpha h_1I_2+h_4I_1,\\
&G_{n}=3h_1\alpha\sum_{k=1}^{n}I_k
I_{n-k}+3h_1\alpha^2\partial_x^{-1}I_{n,y}, \ \ n=2, 3, \cdots.
\end{aligned}$$
The first equation of the conservation law equation (6.26) is
exactly the vc-KP equation (6.1).  Taking the boundary condition of
$p$ into account, the equation (6.26) implies that $I_n's, \ n=1,2,
\cdots$ constitute infinite conserved densities of the vc-KP
equation (6.1). To this end, we remark that as application of these
results, all equations (6.2)-(6.4) are complete integrable under the
constraint $h_2=6h_1e^{\int h_7dt}$, since they possess bilinear
B\"{a}cklund transformation, Lax pair and infinite conservation
laws.
\\[12pt]
{\bf\large  Acknowledgment}

  The work  described in this paper was supported by grants from
   the National Science Foundation of China (No. 10971031),
Shanghai Shuguang Tracking Project (No. 08GG01).

\vspace{1cm}

\end{document}